# Large-Scale Integrated Flexible Tactile Sensor Array for Sensitive Smart Robotic Touch

Zhenxuan Zhao,[1] Jianshi Tang,[1,2*] Jian Yuan,[1] Yijun Li,[1] Yuan Dai,[3*] Jian Yao,[4] Qingtian Zhang,[2] Sanchuan Ding,[2] Tingyu Li,[1] Ruirui Zhang,[3] Yu Zheng,[3] Zhengyou Zhang,[3] Song Qiu,[4] Qingwen Li,[4] Bin Gao,[1,2] Ning Deng,[1,2] He Qian,[1,2] Fei Xing,[2,5] Zheng You,[2,5] Huaqiang Wu[1,2*]

[1]School of Integrated Circuits, Beijing National Research Center for Information Science and Technology (BNRist), Tsinghua University, Beijing, 100084, China.

[2]Beijing Innovation Center for Future Chips (ICFC), Tsinghua University, Beijing, 100084, China.

[3]Tencent Robotics X, Shenzhen, 518000, China.

[4]Key Laboratory of Nanodevices and Applications, Suzhou Institute of Nano-Tech and Nano-Bionics, Chinese Academy of Science, Suzhou, 215123, China.

[5]State Key Laboratory of Precision Measurement Technology and Instruments, Department of Precision Instrument, Tsinghua University, Beijing 100084, China

*Corresponding E-mail: jtang@tsinghua.edu.cn; jessiedai@tencent.com; wuhq@tsinghua.edu.cn

ABSTRACT: In the long pursuit of smart robotics, it has been envisioned to empower robots with human-like senses, especially vision and touch. While tremendous progress has been made in image sensors and computer vision over the past decades, the tactile sense abilities are lagging behind due to the lack of large-scale flexible tactile sensor array with high sensitivity, high spatial resolution, and fast response. In this work, we have demonstrated a 64×64 flexible tactile sensor array with a record-high spatial resolution of 0.9 mm (equivalently 28.2 pixels per inch), by integrating a high-performance piezoresistive film (PRF) with a large-area active matrix of carbon nanotube thin-film transistors. PRF with self-formed microstructures exhibited high pressure-sensitivity of ~385 $kPa^{-1}$ for MWCNTs concentration of 6%, while the 14% one exhibited fast response time of ~3 ms, good linearity, broad detection range beyond 1400 kPa, and excellent cyclability over 3000 cycles. Using this fully integrated tactile sensor array, the footprint maps of an artificial honeybee were clearly identified. Furthermore, we hardware-implemented a smart tactile system by integrating the PRF-based sensor array with a memristor-based computing-in-memory chip to record and recognize handwritten digits and Chinese calligraphy, achieving high classification accuracies of 98.8% and 97.3% in hardware, respectively. The integration of sensor networks with deep learning hardware may enable edge or near-sensor computing with significantly reduced power consumption and latency. Our work could empower the building of large-scale intelligent sensor networks for next-generation smart robotics.

Humans have five basic senses, namely vision, touch, hearing, smell and taste, which play indispensable roles for humans to interact with the environment. From science fictions to real-world applications, it has been long anticipated for robots to feel and act like humans. To achieve that, it is crucial to equip them with a variety of bio-mimicking sensors, which have been an active research and development frontier for decades.[1] Inspiringly, image sensors and computer vision have made tremendous progress in the past years, and can now perform even better than human eyes in many applications.[2] It also largely enriches the sensing abilities of robots with the help of machine learning. By contrast, as a critical sense during physical interactions providing subtle information that vision cannot capture, the developments of tactile sensors and their applications on robots are falling behind. For robots to safely grasp and manipulate objects as humans routinely do, such as opening a bottle without breaking it, high-performance tactile sensors are expected to provide key information, including shape, size, weight, stiffness, *etc*.[3] Therefore, to make robots physically helpful and intelligent, it is highly desired to develop flexible and durable tactile sensors with high sensitivity, high spatial resolution, fast response and large pressure detection range.

In literature, there have been extensive studies on various pressure sensors to mimic the tactile sense of human skin. According to their working mechanisms, flexible pressure sensors can be mainly categorized into three types: piezoelectric,[4-8] capacitive,[9,10] and piezoresistive.[11-13] In general, piezoelectric tactile sensors can have a small size and can operate at high frequency, while capacitive and piezoresistive tactile sensors can have a high pressure sensitivity and are able to measure the applied pressure continuously. Among them, the piezoresistive one is of particular interest to build large-scale tactile sensor networks with high spatial resolution due to its simple structure, convenient signal reading, broad detection range, and low cost.[13] In particular, compositing conductive nanomaterials (*e.g.*, carbon nanotubes,[14, 15] nanofibers,[16] silver nanoparticles,[17] and gold nanowires[18]) with polymer elastomers (*e.g.*, polydimethylsiloxane or PDMS, polyurethane) are widely adopted to synthesize piezoresistive film (PRF) as pressure sensor. In such composites, the piezoresistive effect originated from the tunneling between conductive nanomaterials is largely affected by their geometrical structure (especially the aspect ratio), degree of dispersion and concentration in the composites.[19] In practice, the concentration of conductive nanomaterials is typically quite low (*e.g.*, below 5%) due to the difficulty in uniformly dispersing them in the elastomer precursor and the degree of dispersion is limited by bundle formation during the mixing process. As a result, the pressure sensitivity of PRF is relatively low, especially in the high-pressure range (*e.g.*, over 1000 kPa). It can be enhanced by treating the surface into various microstructures (*e.g.*, pyramid structures,[20, 21] or wrinkled structures[22]), which tend to deform easily under pressure, leading to a sharp increase in the contact area between the PRF and electrodes and hence giving rise to an enhanced pressure sensitivity.[23] However, those microstructures replicated from molds (*e.g.*, silk, paper, patterned silicon wafer, DVD disk, PDMS, and leaf) are usually quite large (typically 15~100 µm),[20, 22, 24-26] which limits the size of an individual pressure sensor and hinders further integration into a large-scale sensor array.

Beyond single-device demonstration for pressure sensor, active matrix with thin-film transistor (TFT) array is usually required for building such large-scale pressure sensor arrays in order to achieve a high spatial resolution and minimize crosstalk between neighboring sensor pixels. For this purpose, a variety of channel materials, including carbon nanotubes (CNTs),[27] semiconductor nanowires,[28] and organic polymers,[29] have been explored to build flexible TFT active matrix by taking advantage of mature micro-fabrication processes.[27, 30] However, their exhibited performances, especially the pressure detection range, sensitivity, and response speed, are still far from satisfactory for the application of smart robotic touch. Beyond sensing, it is further desired to integrate the pressure sensor arrays with signal-processing hardware to enable edge or near-sensor computing with significantly reduced power consumption and latency.[31] For this purpose, artificial intelligence (AI) chips, especially those built with emerging neuromorphic devices like memristors, could process the sensed signals with high efficiency and accuracy by taking advantage of computing-in-memory (CIM) capability.[32, 33] As illustrated in **Figure 1**, such a smart robotic touch system that integrates tactile sensing with CIM is of particular interest for emerging applications such as light-weight robots and Internet of Things (IoTs) where resources are limited.

In this work, we have implemented a tactile system for smart robotic touch by integrating a 64×64 PRF-based sensor array with a memristor-based CIM chip. The system exhibited high accuracies of 98.8% and 97.3% for handwritten digits and Chinese characters recognition in hardware. The PRF was synthesized at low temperature by mixing multi-walled carbon nanotubes (MWCNTs) with thermoplastic polyurethane (TPU) elastomer, in which the self-formed microstructures on the top surface enabled extremely high sensitivity, broad pressure detection range, fast response speed, and excellent cyclability. The self-formed microstructures of the PRFs have a surface roughness of about 8~10 µm, which is much smaller in scale than typical controllable microstructures (*e.g.*, pyramid) reported in literature.[20,21] Therefore, the PRF can be considered quite flat for each pixel with a size of 0.9×0.9 mm$^2$ in the pressure sensor array. By integrating the developed PRF with a 4-inch active matrix of single-walled CNT TFTs, a 64×64 flexible tactile sensor array with ultrahigh spatial resolution was demonstrated and further used to identify the footprint mapping of an artificial honeybee.

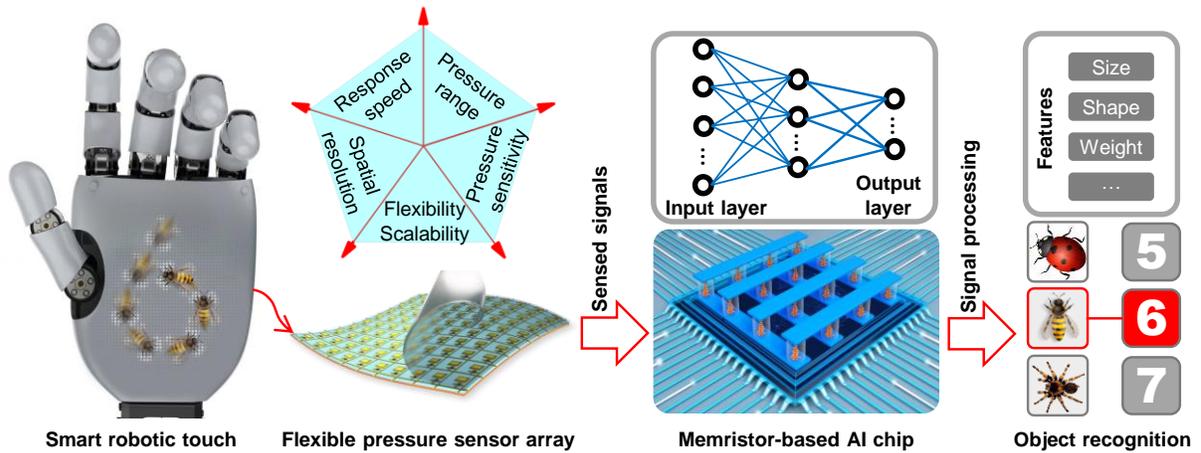

**Figure 1.** Illustration of an exemplary smart robotic touch system. A flexible pressure sensor array is attached on the robotic hand to accurately map out the pressure distribution and mimic the human tactile sense. Besides flexibility, other key metrics of the pressure sensor array include high pressure sensitivity (*e.g.*, >10 kPa$^{-1}$), fast response speed (*e.g.*, <10 ms), high spatial resolution (*e.g.*, <1 mm), large pressure detection range (*e.g.*, >1000 kPa), and good scalability (*e.g.*, array size > 1000). The sensed pressure signals are then fed into memristor-based artificial intelligence (AI) chip that implements an artificial neural network and process the signals with high accuracy and energy efficiency. Important features may be extracted from the sensed information of size, shape, weight, *etc*, and used to recognize objects. Such smart tactile system that combines both sensing and recognition capabilities could endorse future robotics with more functionalities and better intelligence.

RESULTS AND DISCUSSION

**High-performance pressure sensor with PRF.** As discussed above, for composite-based PRFs, realizing homogenous dispersion of high-concentration conductive nanomaterials with high aspect ratio in the elastomer is critical to achieve high sensitivity.[34] In this work, composites of MWCNTs and TPU elastomer were used to synthesize high-performance PRF, because the aspect ratio of MWCNTs, which could be over $10^3$, is among the highest in conductive nanomaterials and they could be wrapped around by TPU uniformly. Also, using a solution-mixing method, N-methyl pyrrolidone (NMP) and N,N-Dimethylformamide (DMF) were chosen as the solvents for MWCNTs and TPU elastomer, respectively, to achieve a high concentration of MWCNTs (up to at least 14%) incorporated into the TPU matrix homogeneously without segregations or MWCNT bundles. Furthermore, the solution mixture was directly cured at a low temperature of 90 °C to promote the formation of self-formed microstructures on the surface of PRF, which could further enhance the pressure sensitivity, especially in the low-pressure regime. The detailed fabrication process is described in the **Experimental Section**. Different MWCNTs weight ratios of 6%, 9%, and 14% were carried out in this study, and the PRF thickness was set to be ~49 μm in consideration of both sensitivity and flexibility.

The structure of the as-fabricated PRF is schematically illustrated in **Figure 2a**, as supported by the following material characterizations. It has two distinct regions: an upper region with a rough surface topography (referred as self-formed microstructures) that has a random arrangement of MWCNTs wrapped with TPU, and a lower region with a flat surface topography that is filled with MWCNTs dispersed uniformly in TPU. The mechanism of realizing such self-formed microstructures of PRF is illustrated and discussed in Figure S1. Considering the weak conductivity of the MWCNTs/TPU composite, helium ion microscope (HIM) and confocal laser scanning microscopy (CLSM) were adopted to investigate the structure of the PRF film. Firstly, to confirm the self-formed microstructures on the top surface, we checked the morphologies and the roughness of the top and bottom surfaces of the fabricated PRF (after peeling it off from the substrate), as shown in Figures 2b-c, respectively. It can be seen MWCNTs dispersed randomly on both surfaces of the PRF without particle bundle or aggregation, reaffirming a homogeneous dispersion of MWCNTs in TPU. The densities of MWCNTs on the top and bottom surfaces are approximately 11.1±3 and 10.5±3 MWCNTs/μm, respectively. The corresponding CLSM results presented the three-dimensional morphology of both surfaces of the PRF. Obviously, the top surface, whose root mean square (RMS) roughness is 8.4 μm, is much rougher than the bottom surface with a RMS roughness of 0.2 μm, while they have a similar density of MWCNTs. Thus, it can be deduced that redundant TPU in the composite precursor agent tended to subside to the bottom of the PRF during the curing process and fill the gaps in between MWCNTs, resulting in a flatter bottom surface. To evaluate the uniformity of the fabricated PRFs, we randomly selected five samples for PRFs with different MWCNTs concentrations to measure the film roughness with 3D CLSM. The results in Figure S2 show good uniformity with roughness of $8.0 \pm 0.36$ μm, $9.9 \pm 0.34$ μm, and $10.6 \pm 0.23$ μm for 6%, 9%, and 14% of MWCNTs, respectively.

Figures 2d-e show the HIM images of the pristine MWCNT from the dispersion solution and the MWCNTs/TPU composite in the PRF, respectively. The brightness contrast in HIM images is related to the absorption and release of $He^+$ ions in the samples, so it could reveal the conductivity difference of materials.[35] From these two images, the outer diameters of the pristine MWCNTs and those in the PRF were measured to be 13±3 nm and 18±3 nm, respectively. The increase of the outer diameter of MWCNTs by ~5 nm is hence attributed to the conformal wrapping of TPU elastomer (poor conductivity) around MWCNTs (high conductivity). The cross section of the PRF film was further investigated in Figure 2f-h, where two distinct regions were clearly visible. The upper region was mainly TPU-wrapped MWCNTs matrix, whereas the lower

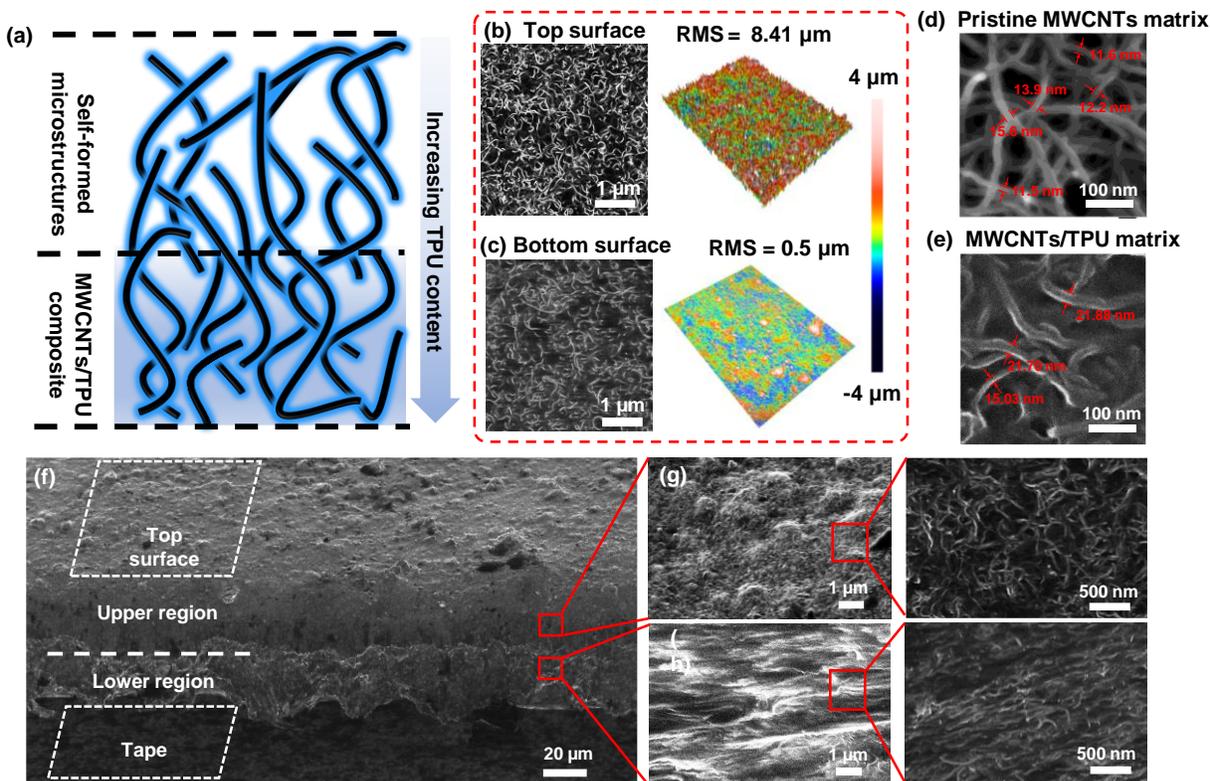

**Figure 2.** Structural characterizations of the MWCNTs/TPU-based PRF. (**a**) Schematic illustration of the structure of the PRF, which has self-formed microstructures in the top region and MWCNTs/TPU composite in the bottom region. (**b**, **c**) HIM images of the (**b**) top and (**c**) bottom surfaces of the PRF with their corresponding three-dimensional CLSM images, respectively. (**d, e**) HIM images of (**d**) the pristine MWCNTs and (**e**) the MWCNTs/TPU matrix in the as-fabricated PRF, respectively. (**f-j**) HIM image of the cross section of the PRF: (**f**) the entire cross section; (**g**) the upper region and its enlarged view; (**h**) the bottom region and its enlarged view.

region was primarily bulk TPU mixed with MWCNTs, resembling the structure illustrated in Figure 2a. This special architecture of PRF could largely enhance its pressure-sensing performance as to be demonstrated in **Figure 3**. **Figure S3** further illustrates the PRF morphology under different pressures. When applying a small pressure, the top region with self-formed microstructures is compressed, leading to the deformation of TPU-wrapped MWCNTs. With increasing pressure, the contact area between the PRF and the top electrode gradually increases. At the same time, the deformation of PRF leads to more current pathways among the TPU-wrapped MWCNTs, resulting in the ultrahigh sensitivity observed in the low-pressure regime.[11, 20, 22, 24-26] Such deformation continues until the gaps in the self-formed microstructures are mostly filled by MWCNTs. After that, as the pressure further increases, the film deformation then mainly comes from the bottom region with MWCNTs/TPU composite, where the compression reduces the distance between adjacent MWCNTs in the composite and hence increases the film conductivity. This trend is expected to continue until the film is extruded into a saturated degree of deformation, where the film conductivity would hardly change with further increasing pressure.

To measure the pressure sensitivity of the PRF, we sandwiched it with two metal electrodes and then measured the current response under different applied pressures in the range of 0~1400 kPa, where the wide pressure range is desired for many robotic applications (*e.g.*, robot dog trotting as to be demonstrated in Figure 3g).[36] The plot of the response current as a function of the applied pressure is provided in **Figure S4**. Figure 3a shows the measured current change of PRFs with three different concentrations (6%, 9%, 14%) of MWCNTs as a function of the applied pressure, in which each data point was the average of five experiments to ensure the data integrity. The pressure sensitivity *S* can be calculated using the following equation:[25, 37, 38]

$$S = \frac{\Delta I/I_0}{P} = \frac{(I-I_0)/I_0}{P} \qquad (1)$$

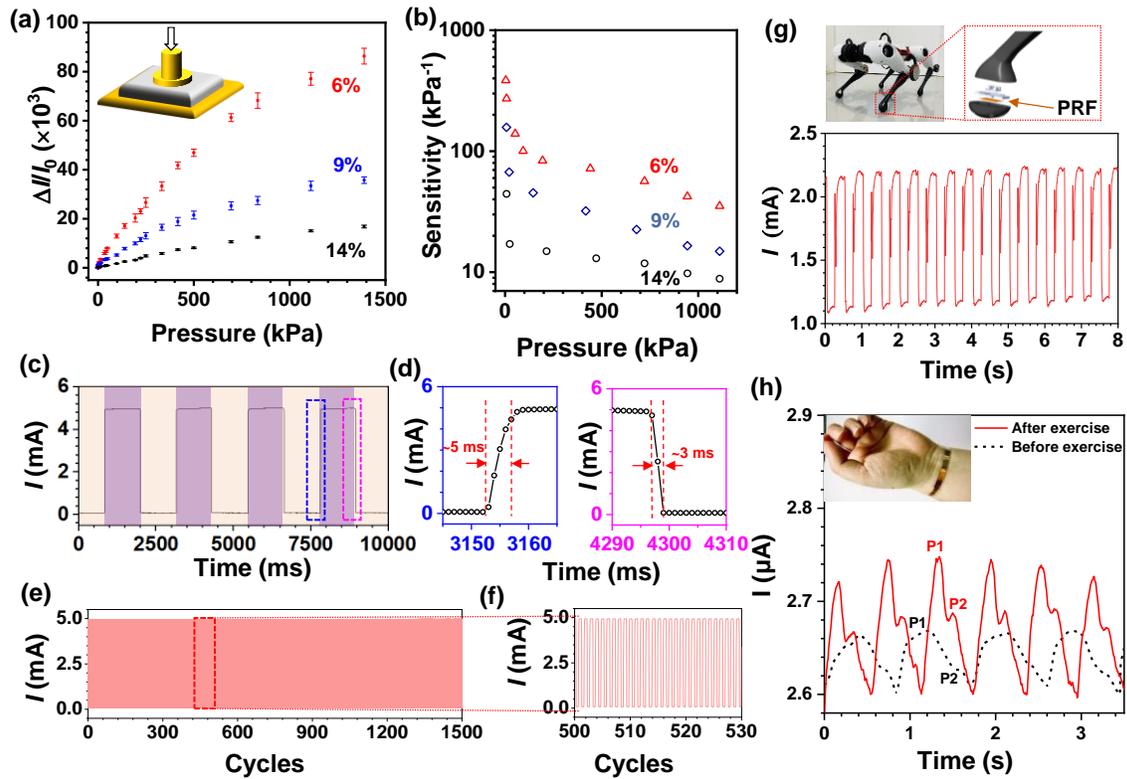

**Figure 3.** Characterizations of the MWCNTs/TPU-based PRF as a pressure sensor. (**a**) Pressure sensing performance of PRFs with three different MWCNTs concentrations of 6%, 9% and 14%. The inset illustrates the experimental setup. (**b**) The extracted pressure sensitivity in the broad range of pressure. (**c**) Application of the PRF pressure sensor to detect the trotting movement of a robot dog. The top panel shows the assembly of the PRF pressure sensor underneath one of the dog legs. (**d**) Enlarged view of the rise (left panel) and fall (right panel) edges in the pressure response test, showing a fast response time of 5 ms and 3 ms, respectively. (**e**) Cyclability test with 1500 cycles, showing no sign of fatigue. (**f**) Enlarged view of multiple cycles in the cyclability test. (**g**) Typical current response of the PRF under repeated pressing test. (**h**)

Demonstration of monitoring human wrist pulses (the inset shows the experimental setup) before and after exercise using the PRF pressure sensor.

Here $I_0$ is the initial current of the PRF under zero pressure, and $I$ is the measured current under the applied pressure $P$. The average $I_0$ are 0.014 µA, 0.200 µA, and 0.625 µA for the PRFs with MWCNTs concentrations of 6%, 9%, and 14%, respectively. It is observed from Figure 3b that the sensitivity $S$ gradually decreases with increasing pressure $P$, and similar behavior has been widely observed in literature-reported PRFs with templated surface structures.[11, 20, 24, 25] The linear fitting of different plots and the log-log ones in the low-pressure range are provided in **Figure S5**. In the low-pressure regime (0~15 kPa), our PRF shows the highest sensitivity that is estimated to be 385 kPa$^{-1}$, 158 kPa$^{-1}$ and 44.4 kPa$^{-1}$ for MWCNTs concentrations of 6%, 9%, and 14% respectively. Here the sensitivity of 385 kPa$^{-1}$ is among the highest reported values measured in the low-pressure regime.[25, 37-49] The difference in the pressure sensitivity for PRFs with increasing MWCNTs concentrations could be attributed to the MWCNTs concentration-dependent elasticity modulus and self-formed microstructures. Future work is still needed to systematically model the deformation of PRF with different MWCNTs concentrations under varying pressures. Besides sensitivity, linearity in pressure sensing and pressure detection range are important metrics for practical applications. A linear pressure response characteristic within a large range helps simplify the readout circuitry and easily estimate the applied pressure from the measured current.[50, 51] Here the PRF with 14% MWCNTs shows the best linearity, and the sensitivity stays roughly constant (~10 kPa$^{-1}$) across a broad pressure range of 15 ~ 1400 kPa, as plotted in Figure 3b. Such a large linear range, along with the broad pressure detection range (1400 kPa), to the best of our knowledge, is the highest reported for pressure sensors so far.[25, 37-49]

Furthermore, the response speed of the PRF pressure sensor was characterized by repeatedly applying and releasing a constant pressure of ~17.68 kPa while recording the current response. The results of four typical cycles of the 14% MWCNTs sample are shown in Figure 3c, where the PRF reproducibly switched between the high-resistance state (HRS) and low-resistance state (LRS) following the applied pressure pattern. The enlarged views of the current response at both rising and falling edges in Figure 3d revealed an ultra-fast response time of about 3~5 ms, much lower than literature-reported values for other PRFs (typically > 10 ms) as summarized in **Table S1**. It also indicates that our pressure sensor can respond nearly ten times faster than the reaction time of human skin.[52] In addition, Figure 3e shows the cyclability test with 1500 cycles and Figure 3f shows the enlarged view of multiple cycles in the middle of the test. Cyclability test up to 3000 cycles on another PRF sample is shown in **Figure S6**. We can see well-repeated switching between HRS and LRS without much degradation, confirming that our PRF is highly durable and robust.

By developing PRFs with different MWCNTs concentrations, the varying pressure sensing properties allow us to select suitable PRF films to meet the demand of specific applications. To demonstrate the application of such high-performance PRF, we firstly assembled a PRF sensor with 14% MWCNTs, thanks to its wide pressure detection range and fast response, and placed it underneath the foot of a robot dog to detect its movement during trotting (see **Video S1**), where the maximum pressure applied is about 1000 kPa. In this experiment, the assembly and testing method of the PRF pressure sensor is described in **Figure S7.** The measured current response from multiple cycles shown in Figure 3g. The demonstration certifies the broad dynamic range, fast response and excellent reliability of the developed PRF pressure sensor.

A human pulse monitoring experiment was carried out with another PRF pressure sensor with 6% MWCNTs to take the advantage of its ultrahigh sensitivity as shown in Figure 3h. An increasing heart rate and intensity from ~70 beats/min before exercise to ~101 beats/min after exercise was detected. The complete pulse shape with two characteristic peaks marked as P1 and P2 was also clearly identified,[37, 44, 48] showing excellent sensitivity and fast response. All the above results demonstrate the outstanding pressure sensing properties of the developed PRF in terms of high sensitivity, broad pressure detection range, good linearity, fast response time, and superb cyclability, making it an excellent candidate for high-performance tactile sensors. Besides normal force detection, the PRF sensors could be assembled into an array to detect the direction of the applied force by calculating the applied pressure on each sensor in the array.[53]

**High spatial resolution flexible pressure sensor array.** For robotic touch, a large-scale integrated pressure sensor array with high spatial resolution, instead of a single device or multiple discrete sensors, is usually required to detect the location, shape, size, and pressure distribution of the incoming contact with the environment on the robot. Here we integrated our low-temperature processed PRF with a 64×64 active matrix made with single-walled CNT TFTs to build a large-scale integrated pressure sensor array covering 4-inch area. The active matrix was first fabricated on a flexible polyimide (PI) film using a 4-inch silicon handling wafer. This process yielded an ultra-flat and clean PI surface with a roughness of less than 1 nm (**Figure S8**), which provided a flatter and more convenient substrate than the free-standing PI film used in our previous work for the subsequent fabrication of CNT-TFT active matrix. The detailed fabrication process is described in the **Experimental Section** (see **Figure S9**). In brief, the TFT has a local bottom-gate structure with 2 nm Ti/40 nm Pd/1 nm Ti as the gate, and 40 nm $Al_2O_3$/10 nm $HfO_2$ as the gate oxide. The channel was made of a high-density CNT random network deposited by standard lift-coating method using a polymer-sorted CNT solution (semiconducting purity >99.99%).[54, 55] Compared to our previous work,[27] the scale of the active matrix was increased by 16 times, from 16×16 to 64×64, leading to a much-enhanced spatial resolution of 0.9 mm that corresponded to 28.2 pixels per inch (ppi) or 123 transistors/cm$^2$. Both numbers are the highest reported so far for active matrix-based piezoresistive pressure sensor array[27, 30, 56-61] as compared in **Figure S10**. This spatial resolution is close to the reported spatial acuity of pattern-sensing capability of about 0.5 mm on human skin.[62]

The images and schematic of the fabricated CNT TFT active matrix are shown in **Figure 4a-c**. Figure 4d is the scanning electron microscope (SEM) image of the high-density CNT thin film in the channel, which has a dimension of 8 μm in length and 100 μm in width. Figure 4e shows the photograph of the active matrix after peeling off from the handling substrate. Electrical measurements were done to characterize the CNT TFT performance in the active matrix. Figure 4f shows the $I_{ds}$-$V_{gs}$ transfer characteristics from a typical CNT TFT, which exhibited an on-state current of $I_{on}$ = 38.7 μA at $V_{ds}$ of -1 V, a subthreshold slope of $SS$ = 288 mV/dec, and a high on/off ratio above $10^6$ owing to the high purity of the semiconducting CNT solution. The inset of Figure 4g plots the color mapping of $I_{on}$ (mean value of 35.8 μA) from the entire active matrix. 4083 out of the 4096 pixels were found to be well functional, indicating a high yield of 99.7% for a large-area sensor array fabricated in a university lab. Based on the statistical data measured from over 4000 TFTs, the carrier mobility was estimated to be $\mu$ = 8.2±4.9 cm$^2$/V·s as shown in the histogram in Figure 4g. In addition, repeated bending tests on the active matrix after peeling off from the handling wafer were carried out. The CNT TFT made on PI substrate exhibited excellent flexibility and durability under different bending radii down to 10 mm (**Figure S11**).

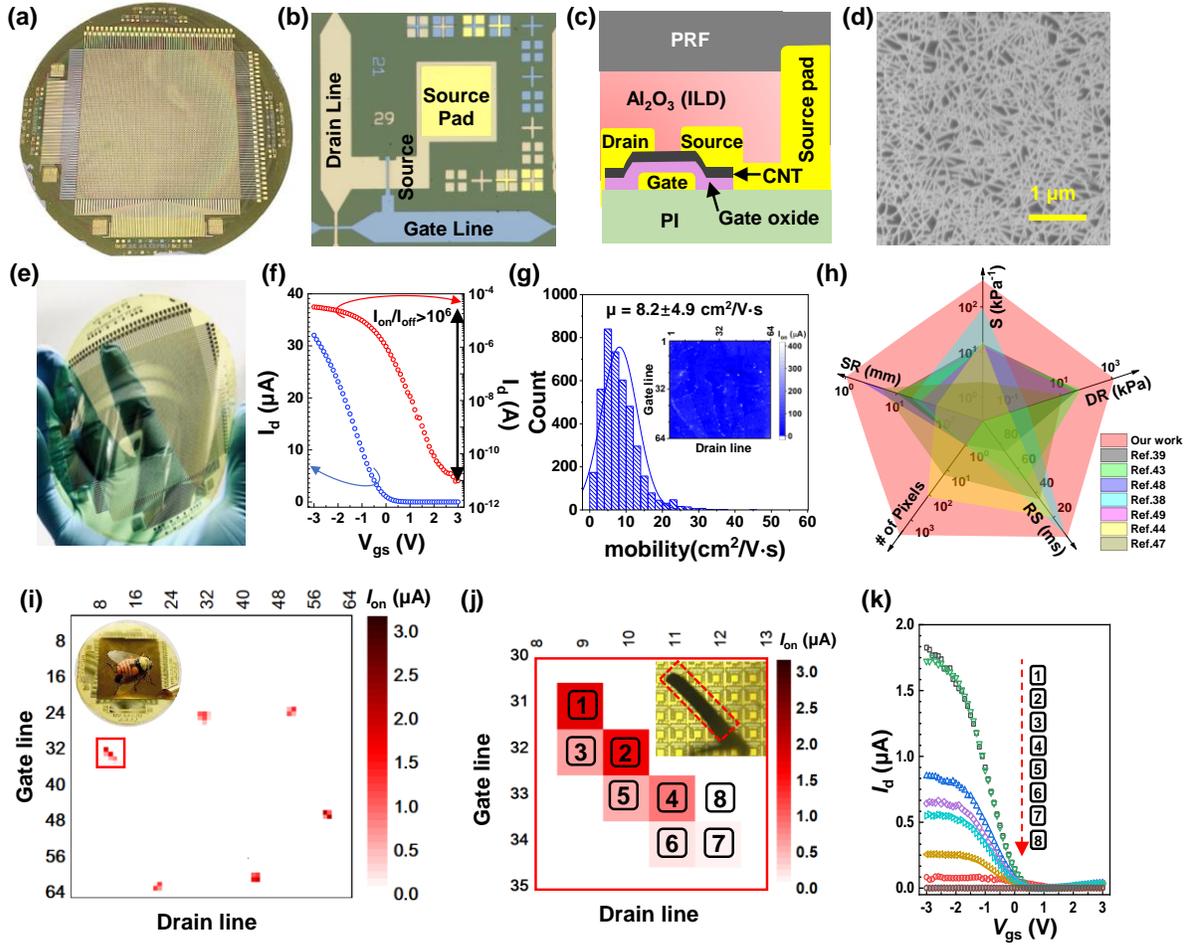

**Figure 4.** Large-scale integrated flexible pressure sensor array with CNT TFT active matrix. (**a**) Photograph of the fabricated active matrix with 64×64 CNT TFTs before peeling off from the silicon handling wafer. (**b**) Microscope image of one typical pixel of the active matrix. The TFT channel with a dimension of $W_{ch}/L_{ch}$ = 100 μm/8 μm. (**c**) The cross-section schematic of the CNT TFT as one pixel in the active matrix. (**d**) Scanning electron microscope (SEM) image of the deposited high-density CNT thin film. (**e**) Photograph of the fabricated active matrix after peeling off from the handling wafer. (**f**) $I_{ds}$-$V_{gs}$ transfer curve of a typical CNT TFT at $V_{ds}$= -1 V in linear (left axis) and logarithmic (right axis) scales. (**g**) Histogram of the extracted mobility from over 4000 CNT TFTs. The inset shows the color map of the on-state current $I_{on}$ for the active matrix. (**h**) Comparison of the five key metrics of literature-reported flexible pressure sensors: pressure sensitivity (S), detection range (DR), spatial resolution (SR), scalability (number of pixels), and response speed (RS). Our work represents state-of-the-art performance for flexible pressure sensors. (**i**) Current mapping for the pressure sensing of an artificial honeybee (six feet) standing on the pressure sensor array. The inset shows the photograph. (**j**) Enlarged current mapping near the honeybee's hindfoot. The inset shows the microscope image. (**k**) Transfer curves of the eight CNT TFTs underneath the honeybee's hindfoot.

By applying a large-size PRF (with 14% MWCNTs) on top of the fabricated active matrix, we hereby built a fully integrated sensor array and further investigated its pressure sensing performance (**Figure S12**). An individual pixel was first tested by applying different pressures on the PRF while measuring the transfer characteristics of the CNT TFT at $V_{ds}$ = -1 V, and the results are shown in Figure S12b. It is observed that the TFT current $I_{ds}$ gradually increased with the applied pressure rising from 0 to ~118 kPa, showing a large output current dynamic range of over 1000× for pressure sensing. The extracted pressure sensitivity for the fully integrated pressure sensor was about 42.5~12.9 kPa$^{-1}$ as shown in Figure S12c, which is close to the value obtained from PRF sensor itself presented in Figure S13. In addition, the pressure response time of the sensor array was also measured by repeatedly applying and releasing a constant pressure of ~15.7 kPa while biasing the CNT TFT at $V_{ds}$ = -1 V and $V_{gs}$ = -3 V. As shown in Figure S12d, the integrated sensor array exhibited an ultra-fast response time of ~4 ms, which is consistent with the values measured with PRF alone in Figure 3e. The excellent flexibility of the flexible CNT TFT active matrix was also verified in the bending test in **Figure S14**. Overall, Figure 4h compares the five key metrics, namely pressure sensitivity, detection range, spatial resolution, number of pixels, and response speed, of flexible pressure sensors reported in literature.[38-40, 42, 44,]

[45, 49] It is evident that our work demonstrates the state-of-the-art performance for flexible pressure sensors.

Furthermore, such a large-scale integrated pressure sensor array with high sensitivity and high spatial resolution can be used to accurately map out the pressure distribution over a large area, which may help identify the shape and weight of small objects to fully mimic human tactile sense.[13] As an example, here we demonstrated the identification of the footprint map of an artificial honeybee (weight of 6.7 grams and feet of ~0.55 mm in width) standing on the sensor array, as shown in Figure 4i. The locations of its six feet were correctly identified in the current mapping of the pressure sensor array. Zooming into one of its feet, the left hindfoot as an example, we can see that eight sensor pixels in the array were pressed with different pressures applied on each of them (Figure 4j). The stripe pattern matched well with the shape and size of the honeybee's foot. The complete transfer curves of the eight pressed pixels are shown in Figure 4k. The variation of $I_{on}$ (measured at $V_{ds}$ = -1 V and $V_{gs}$ = -3 V) in the range of 1.83~0.084 µA reflected the landing force from the hindfoot increased from the top to bottom. Besides this standing still position, we also measured the current mapping when the honeybee was standing on its three feet during clawing, as shown in **Figure S15**. It is noted that $I_{on}$ increased as a sign of increasing pressure due to the reduced area to support the entire weight of the honeybee. These results demonstrate the great potential of using the integrated PRF-based sensor array as an electronic tactile system for smart robotic touch.

**Smart tactile system for sensing and recognition.** Besides identifying the pressure map associated with the shape of objects, the collected sensor data could also help recognize different patterns represented in the data. It is then motivated to further integrate the sensor array with AI hardware and take advantage of deep learning algorithms for efficient data processing, enabling

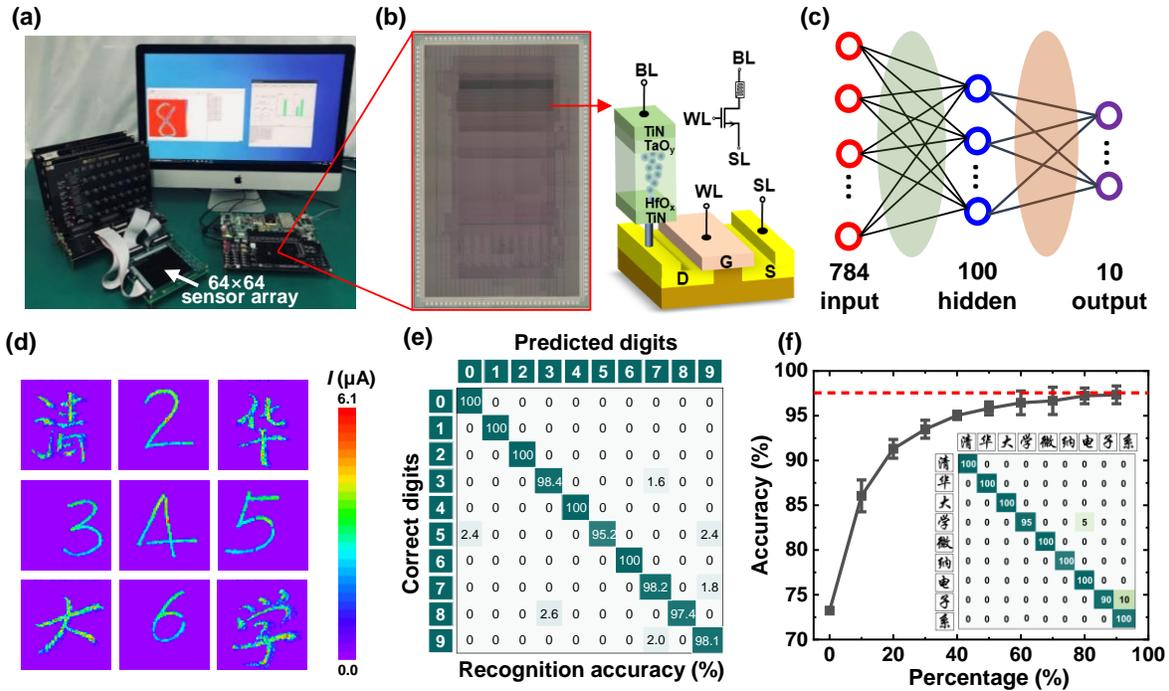

**Figure 5.** Prototype of smart tactile system that combines sensing and recognition. (**a**) Photograph of the prototype smart tactile system integrating pressure sensor array for signal sensing with memristor-based computing-in-memory (CIM) chip for signal processing (recognition). The arrow indicates the 64×64 PRF-based pressure sensor array on the customized PCB board. (**b**) Die photograph of the fully integrated CIM chip with ~160k memristors and peripheral circuits. The inset shows the schematic of one-transistor-one-resistor (1T1R) unit cell with TiN/HfO$_x$/TaO$_y$/TiN memristor in the CIM chip. (**c**) The implemented multi-layer perceptron (MLP) of 784×100×10. (**d**) Typical images of the recorded handwritten digits and Chinese characters for recognition. (**e**) Confusion matrix of the recognition accuracy for the collected 3099 images of handwritten digits. High accuracies of 99.2% and 98.8% were achieved for training and testing, respectively. (**f**) Recognition accuracy of the testing set as a function of the percentage of the recorded 900 images of handwritten Chinese characters used for the training set. Each experiment was repeated for five times using randomly assigned testing set.

The experimentally achieved recognition accuracy with our CIM chip is equivalent to the software baseline of 97.6±1.2%. The inset shows the typical confusion matrix with recognition accuracy of 98.3%.

future edge or near-sensor computing with significantly reduced power consumption and latency.[31] As a proof-of-concept demonstration, we built a prototype tactile hardware system for smart robotic touch by integrating the 64×64 PRF-based sensor array with a memristor-based CIM chip to recognize handwritten digits (**Figure 5a**). As an emerging nonvolatile memory, analog resistive switching memory (RRAM), also known as memristor, has been extensively studied as a promising neuromorphic device to implement artificial neural networks with high energy efficiency. As reported in our previous work,[63] the fully integrated CIM chip was fabricated in a commercial foundry using a standard 130nm CMOS process (Figure 5b), and it implemented a multi-layer perceptron (MLP) of 784×100×10 using ~160k memristors (Figure 5c). This chip could achieve a peak energy efficiency of 78.4 TOPS/W and a fast inference speed of 77 μs/image for the Modified National Institute of Standards and Technology (MNIST) dataset.

For the convenience of collecting sufficient amount of data for the artificial neural network training and inference, here we built a customized printed circuit board (PCB) with 64×64 PRF-based sensor arrays (Figure 5a) and a field-programmable gate array (FPGA) evaluation board (Zynq-7015) with peripheral circuits for signal readout. The currents of all 4096 pressure sensors were read sequentially without cross-talk. The read time was 5 μs for each cell, and the entire sensor array was sampled at a frame rate of 50 fps. Alternative array design in literature can be adopted to reduce the number of sample channels.[64] All the frames recorded during the writing of one digit (duration of ~1 s) were combined to produce one image of 64×64 pixels (Figure 5d). In total 3099 images of handwritten digits were collected, from which 2598 images were randomly selected as the training set and the rest 501 images were used as the testing set. The images of handwritten digits were down-sampled to 28×28 pixels and then fed into the fully integrated memristor CIM chip for classification. As shown in Figure 5e, high accuracies of 99.2% and 98.8% were achieved for training and testing, respectively, which are comparable to the values achieved in software simulations. Beyond handwritten digits, more complex patterns such as Chinese characters were collected in a similar fashion using the pressure sensor array for classification. Here we chose 9 Chinese characters that represent our department affiliation at Tsinghua University, and collected a dataset of 900 images (100 for each character) by writing on the 64×64 sensor array. Different percentages from 0% to 90% of the handwritten dataset were added to the training set, which consisted of 12150 images (1350 for each character) generated by computer, for the neural network training. The rest of the handwritten datasets were used as the testing set, and the recognition accuracy is shown in Figure 5f. For statistical analysis, each experiment was repeated five times using a randomly assigned testing set. It can be seen that the accuracy achieved with our CIM chip increased significantly from 73.2% to 97.3±1.0% when 90% of the sensor-generated handwritten dataset was included for training, suggesting the smart tactile system became more and more intelligent to recognize personal handwriting. Such experimentally achieved accuracy is almost equivalent to the software baseline of 97.6±1.2% obtained by computer with numerical simulations. These results not only indicated the high quality of pressure mapping images collected by the PRF-based pressure sensor array but also confirmed the feasibility of fully integrated tactile system combining both sensing and recognition for smart robotic touch application.

CONCLUSION

In summary, we have demonstrated a state-of-the-art large-scale integrated flexible pressure sensor array built with a 4-inch size active matrix of 64×64 CNT TFTs and high-performance PRF. By homogeneously mixing MWCNTs with TPU elastomer, the low-temperature processed PRF exhibited extremely high sensitivity (up to 385 kPa$^{-1}$), broad detection range (0~1400 kPa), fast response speed (3~5 ms), and outstanding cyclability over 3000 cycles. Detailed structural analysis with HIM and CLSM suggested the excellent pressure sensing performance can be attributed to the self-formed microstructures on the top surface of the PRF. With a high yield of 99.7% for the active matrix, the fully integrated tactile sensor array featured with a record-high scale of 4096 pixels and spatial resolution of 0.9 mm along with high sensitivity and fast response speed. Accurate detections of honeybee footprint map and human wrist pulses have been demonstrated. Furthermore, a prototype smart tactile system possessing both signal sensing and recognition capabilities has been implemented by integrating the 64×64 pressure sensor array with a memristor-based CIM chip. Using this hardware system, software-equivalent classification accuracies of 98.8% and 97.3% have been achieved in experiments for handwritten digits and Chinese characters recognition, respectively. Our work highlights the tremendous potential of using the developed tactile sensor array for edge computing or near-sensor computing to endorse future robotics with more functionalities and better intelligence.

EXPERIMENTAL SECTION

**Fabrication and characterization of MWCNTs/TPU composite as PRF.** Materials: MWCNTs dispersion in N-methyl pyrrolidone (NMP) was purchased from Chinese Academy of Sciences with an outer diameter of 8~15 nm, an inner diameter of 3~5 nm, length of ~50 μm, and a weight ratio of 2.1%; TPU was purchased from Baoruilong Polymer Material (Tianjin) Company; N,N-Dimethylformamide solution (DMF) was purchased from China National

Pharmaceutical Group Corporation. Firstly, TPU was dispersed into DMF solution and stirred thoroughly until forming a transparent liquid at room temperature. The weight ratio of TPU:DMF was set to be 1:5. Then, after ultrasonic pretreatment, the MWCNTs dispersion was added into the above mixture, which was then stirred uniformly to obtain the PRF precursor agent. The added quantity of the MWCNTs dispersion was calculated according to the total weight ratio of MWCNTs in the PRF. In this work, weight ratios of 6%, 9%, and 14% were used. The PRF was prepared by casting the precursor agent on the target substrate and it can be cured directly when baking at 90 ºC for 4 hours. The morphology and surface roughness of the fabricated PRF was tested using a helium-ion microscope (HIM, Zeiss ORION NanoFab Helium Ion Microscope) and a confocal laser scanning microscope (CLSM, Lycra #DCM8). Electrical characterizations of the PRF were done on a customized probe station with Keithley 2400 source meter. It should be noted that the DMF solvent plays a critical role in the preparation of high-quality PRF and its pressure sensing performance. A comparison experiment using different solvents was performed and the results are shown in **Figures S16** and **S17**.

**Fabrication of the CNT TFT-based active matrix.** The fabrication process of the CNT TFT-based active matrix is illustrated in **Figure S9**. To prepare the flexible substrate, polyimide (PI) curing agent was uniformly spin coated on a 4-inch silicon handling wafer. It was then cured under increasing temperature from 50 ºC to 250 ºC and then maintained at 250 ºC for 2 hours. After that, local bottom gates along with 64 gate lines were first patterned on PI by standard photolithography and metal deposition of 2 nm titanium (Ti)/40 nm palladium (Pd)/1 nm Ti. Then, atomic layer deposition (ALD) of 40 nm $Al_2O_3$/10 nm $HfO_2$ at 200 °C was done to grow the gate dielectric. According to the capacitance measurement of the same stack of ALD oxides deposited on a control sample, the equivalent relative dielectric constant was calculated to be $\varepsilon_r =$ 7.36. High-density CNT random network was deposited as the channel by the standard lift-coating method using a highly purified CNT solution (purity >99.99%) that was prepared in a similar fashion as our previous work.[54,55] Source/drain contacts, along with 64 drain lines were then defined by another photolithography step and e-beam evaporation of 0.2 nm Ti/60 nm Pd/2 nm Ti. Then, CNT TFT channels with the length of $L_{ch}$ = 8 μm and width of $W_{ch}$ = 100 μm were defined using photolithography followed by oxygen plasma etching. To protect the CNT TFTs and also adjust the threshold voltage, a 20 nm-thick $Al_2O_3$ passivation layer was deposited by ALD at 150 °C, which also served as the interlayer dielectric (ILD). Via contacts were made on each source pad through pad opening lithography and etching of the ILD layer. Then contact pads made of 5 nm Ti/100 nm Au were fabricated by standard e-beam evaporation and lift-off process. After that, the prepared MWCNTs/TPU precursor agent was used to make the PRF with 14% MWCNTs covering the entire active matrix. Finally, the fully integrated flexible tactile pressure sensor array can be delaminated from the silicon substrate for electrical measurements.

**Fabrication of memristor-based CIM chip.** The full system integrated memristor-based CIM chip was fabricated in a commercial foundry using a standard 130nm CMOS process to implement a multi-layer perceptron (MLP) of 784×100×10.[63] In this chip, ~160K TiN/$HfO_x$/$TaO_y$/TiN filamentary-type memristors were monolithically integrated with all the peripheral circuits including analog-to-digital converters (ADCs), digital-to-analog converters (DACs), drivers and buffers. A differential pair of one-transistor-one-resistor (1T1R) unit cell was used to represent one synaptic weight. The size of the memristor was 0.5μm×0.5μm. The $HfO_x$ resistive switching layer was about 8nm and the $TaO_x$ thermal enhanced layer was about 45nm. The top and bottom electrodes of the memristor were both 30nm-thick TiN deposited by sputtering.

ACKNOWLEDGMENT

This work was supported in part by Tencent Robotics X Lab.

Author Contributions

J.T. and Y.D. conceived and designed the experiments. Z.Z. synthesized the piezoresistive film. J.Y., Y.L. and T.Y. fabricated the active matrix. J.Y., S.Q. and Q.L. provided the carbon nanotubes. Z.Z. and J.Y. performed the measurements with contributions from S.D., B.G., N.D., H.Q., F.X., and Z.Y.. Y.D., R.Z., Y.Z., and Z.Z. contributed to the characterizations of the piezoresistive film. Q.Z. performed the neural network simulations. Z.Z., J.T., and Y.D. wrote the paper. All authors discussed the results and commented on the manuscript. J.T., Y.D., and H.W. supervised the project.

Funding Sources

Tsinghua University - Tencent Joint Laboratory for Internet Innovation Technology JR201985; Tsinghua University - Tencent Joint Laboratory for Internet Innovation Technology JR2021TEG002.

Notes

The authors declare no competing financial interest.

ASSOCIATED CONTENT

**Supporting Information**.

The Supporting Information is available free of charge.

Characteristics of recently reported PRFs, illustration of the forming process of self-formed microstructures, roughness of PRFs with different MWCNT concentrations, illustration of the deformation of PRF with increasing pressure, additional data on the pressure response of PRFs, cyclability test of another PRF sample with 3000 cycles, illustration of the assembly of the PRF pressure sensor on a robot dog's foot, AFM image of the spin-coated polyimide film, illustration of the fabrication processes of the flexible pressure sensor array, comparison of literature-reported active matrix-based flexible pressure sensors, bending test on the CNT active matrix, additional pressure-sensing data on the fully integrated flexible pressure sensor array, pressure sensing performance of the 14% PRF in the low pressure range, bending test of the integrated flexible pressure sensory array, additional data on the honeybee footprint mapping, comparison of PRFs synthesized using different solvents (PDF), and video of robot dog trotting with PRF pressure sensor (MP4).

# REFERENCES

1. Alatise, M. B.; Hancke, G. P., A Review on Challenges of Autonomous Mobile Robot and Sensor Fusion Methods. *IEEE Access* **2020,** *8*, 39830-39846.
2. Pauli, J., *Learning-Based Robot Vision*. Springer: Berlin, Heidelberg, 2001; Vol. 2048.
3. Kappassov, Z.; Corrales, J.-A.; Perdereau, V., Tactile sensing in dexterous robot hands - Review. *Robotics and Autonomous Systems* **2015,** *74*, 195-220.
4. Zhang, C.; Chen, J.; Xuan, W.; Huang, S.; You, B.; Li, W.; Sun, L.; Jin, H.; Wang, X.; Dong, S.; Luo, J.; Flewitt, A. J.; Wang, Z. L., Conjunction of Triboelectric Nanogenerator with Induction Coils As Wireless Power Sources and Self-Powered Wireless Sensors. *Nature communications* **2020,** *11* (1), 58-58.
5. Pu, X.; Liu, M.; Chen, X.; Sun, J.; Du, C.; Zhang, Y.; Zhai, J.; Hu, W.; Wang, Z. L., Ultrastretchable, Transparent Triboelectric Nanogenerator as Electronic Skin for Biomechanical Energy Harvesting and Tactile Sensing. *Science Advances* **2017,** *3* (5), e1700015.
6. Zhang, Z.; Liu, S.; Pan, Q.; Hong, Y.; Shan, Y.; Peng, Z.; Xu, X.; Liu, B.; Chai, Y.; Yang, Z., Van Der Waals Exfoliation Processed Biopiezoelectric Submucosa Ultrathin Films. *Advanced Materials* **2022,** *34* (26), 2200864.
7. Liu, S.; Shan, Y.; Hong, Y.; Jin, Y.; Lin, W.; Zhang, Z.; Xu, X.; Wang, Z.; Yang, Z., 3D Conformal Fabrication of Piezoceramic Films. *Advanced Science* **2022,** *9* (18), 2106030.
8. Zou, D.; Liu, S.; Zhang, C.; Hong, Y.; Zhang, G.; Yang, Z., Flexible and Translucent PZT Films Enhanced by the Compositionally Graded Heterostructure for Human Body Monitoring. *Nano Energy* **2021,** *85*, 105984.
9. Kaisti, M.; Panula, T.; Leppanen, J.; Punkkinen, R.; Tadi, M. J.; Vasankari, T.; Jaakkola, S.; Kiviniemi, T.; Airaksinen, J.; Kostiainen, P.; Meriheina, U.; Koivisto, T.; Pankaala, M., Clinical Assessment of a Non-Invasive Wearable MEMS Pressure Sensor Array for Monitoring of Arterial Pulse Waveform, Heart Rate and Detection of Atrial Fibrillation. *Npj Digital Medicine* **2019,** *2*, 39.
10. Bai, N.; Wang, L.; Wang, Q.; Deng, J.; Wang, Y.; Lu, P.; Huang, J.; Li, G.; Zhang, Y.; Yang, J.; Xie, K.; Zhao, X.; Guo, C. F., Graded Intrafillable Architecture-Based Iontronic Pressure Sensor with Ultra-Broad-Range High Sensitivity. *Nature communications* **2020,** *11* (1), 209-209.
11. Wang, L.; Peng, H.; Wang, X.; Chen, X.; Yang, C.; Yang, B.; Liu, J., PDMS/MWCNT-Based Tactile Sensor Array with Coplanar Electrodes for Crosstalk Suppression. *Microsystems & Nanoengineering* **2016,** *2*, 16065.
12. Ma, Y.; Liu, N.; Li, L.; Hu, X.; Zou, Z.; Wang, J.; Luo, S.; Gao, Y., A Highly Flexible and Sensitive Piezoresistive Sensor Based on Mxene with Greatly Changed Interlayer Distances. *Nature Communications* **2017,** *8*, 1207.
13. Sundaram, S.; Kellnhofer, P.; Li, Y.; Zhu, J.-Y.; Torralba, A.; Matusik, W., Learning the Signatures of the Human Grasp Using a Scalable Tactile Glove. *Nature* **2019,** *569* (7758), 698-702.
14. Yamada, T.; Hayamizu, Y.; Yamamoto, Y.; Yomogida, Y.; Izadi-Najafabadi, A.; Futaba, D. N.; Hata, K., A Stretchable Carbon Nanotube Strain Sensor for Human-Motion Detection. *Nature Nanotechnology* **2011,** *6* (5), 296-301.
15. Habibi, M.; Darbari, S.; Rajabali, S.; Ahmadi, V., Fabrication of a Graphene-Based Pressure Sensor by Utilising Field Emission Behavior of Carbon Nanotubes. *Carbon* **2016,** *96*, 259-267.
16. Pang, C.; Lee, G.-Y.; Kim, T.-i.; Kim, S. M.; Kim, H. N.; Ahn, S.-H.; Suh, K.-Y., A Flexible and Highly Sensitive Strain-Gauge Sensor Using Reversible Interlocking of Nanofibres. *Nature Materials* **2012,** *11* (9), 795-801.
17. Matsuhisa, N.; Inoue, D.; Zalar, P.; Jin, H.; Matsuba, Y.; Itoh, A.; Yokota, T.; Hashizume, D.; Someya, T., Printable Elastic Conductors by *In Situ* Formation of Silver Nanoparticles from Silver Flakes. *Nature Materials* **2017,** *16* (8), 834-840.
18. Gong, S.; Schwalb, W.; Wang, Y.; Chen, Y.; Tang, Y.; Si, J.; Shirinzadeh, B.; Cheng, W., A Wearable and Highly Sensitive Pressure Sensor with Ultrathin Gold Nanowires. *Nature Communications* **2014,** *5*, 3132.
19. Hu, N.; Karube, Y.; Arai, M.; Watanabe, T.; Yan, C.; Li, Y.; Liu, Y.; Fukunaga, H., Investigation on Sensitivity of a Polymer/Carbon Nanotube Composite Strain Sensor. *Carbon* **2010,** *48* (3), 680-687.
20. Jung, M.; Vishwanath, S. K.; Kim, J.; Ko, D.-K.; Park, M.-J.; Lim, S.-C.; Jeon, S., Transparent and Flexible Mayan-Pyramid-Based Pressure Sensor Using Facile-Transferred Indium tin Oxide for Bimodal Sensor Applications. *Scientific Reports* **2019,** *9*, 14040.
21. Ge, J.; Wang, X.; Drack, M.; Volkov, O.; Liang, M.; Bermudez, G. S. C.; Illing, R.; Wang, C.; Zhou,

S.; Fassbender, J.; Kaltenbrunner, M.; Makarov, D., A Bimodal Soft Electronic Skin for Tactile and Touchless Interaction in Real Time. *Nature Communications* **2019,** *10*.
22. Zeng, X.; Wang, Z.; Zhang, H.; Yang, W.; Xiang, L.; Zhao, Z.; Peng, L.-M.; Hu, Y., Tunable, Ultrasensitive, and Flexible Pressure Sensors Based on Wrinkled Microstructures for Electronic Skins. *ACS Applied Materials & Interfaces* **2019,** *11* (23), 21218-21226.
23. Choong, C.-L.; Shim, M.-B.; Lee, B.-S.; Jeon, S.; Ko, D.-S.; Kang, T.-H.; Bae, J.; Lee, S. H.; Byun, K.-E.; Im, J.; Jeong, Y. J.; Park, C. E.; Park, J.-J.; Chung, U. I., Highly Stretchable Resistive Pressure Sensors Using a Conductive Elastomeric Composite on a Micropyramid Array. *Advanced Materials* **2014,** *26* (21), 3451-3458.
24. Wang, X.; Gu, Y.; Xiong, Z.; Cui, Z.; Zhang, T., Silk-Molded Flexible, Ultrasensitive, and Highly Stable Electronic Skin for Monitoring Human Physiological Signals. *Advanced Materials* **2014,** *26* (9), 1336-1342.
25. Tian, H.; Shu, Y.; Wang, X.-F.; Mohammad, M. A.; Bie, Z.; Xie, Q.-Y.; Li, C.; Mi, W.-T.; Yang, Y.; Ren, T.-L., A Graphene-Based Resistive Pressure Sensor with Record-High Sensitivity in a Wide Pressure Range. *Scientific Reports* **2015,** *5*, 8603.
26. Jian, M.; Xia, K.; Wang, Q.; Yin, Z.; Wang, H.; Wang, C.; Xie, H.; Zhang, M.; Zhang, Y., Flexible and Highly Sensitive Pressure Sensors Based on Bionic Hierarchical Structures. *Advanced Functional Materials* **2017,** *27* (9), 1606066.
27. Nela, L.; Tang, J.; Cao, Q.; Tulevski, G.; Han, S.-J., Large-Area High-Performance Flexible Pressure Sensor with Carbon Nanotube Active Matrix for Electronic Skin. *Nano Letters* **2018,** *18* (3), 2054-2059.
28. Duan, X. F.; Niu, C. M.; Sahi, V.; Chen, J.; Parce, J. W.; Empedocles, S.; Goldman, J. L., High-Performance Thin-Film Transistors Using Semiconductor Nanowires and Nanoribbons. *Nature* **2003,** *425* (6955), 274-278.
29. Lee, H.-S.; Lee, J. S.; Jung, A. R.; Cha, W.; Kim, H.; Son, H. J.; Cho, J. H.; Kim, B., Processing Temperature Control of a Diketopyrrolopyrrole-Alt-Thieno 2,3-B Thiophene Polymer for High-Mobility Thin-Film Transistors and Polymer Solar Cells with High Open-Circuit Voltages. *Polymer* **2016,** *105*, 79-87.
30. Wang, S.; Xu, J.; Wang, W.; Wang, G.-J. N.; Rastak, R.; Molina-Lopez, F.; Chung, J. W.; Niu, S.; Feig, V. R.; Lopez, J.; Lei, T.; Kwon, S.-K.; Kim, Y.; Foudeh, A. M.; Ehrlich, A.; Gasperini, A.; Yun, Y.; Murmann, B.; Tok, J. B. H.; Bao, Z., Skin Electronics from Scalable Fabrication of an Intrinsically Stretchable Transistor Array. *Nature* **2018,** *555* (7694), 83-88.
31. Zhou, F.; Chai, Y., Near-Sensor and In-Sensor Computing. *Nature Electronics* **2020,** *3* (11), 664-671.
32. Yao, P.; Wu, H.; Gao, B.; Tang, J.; Zhang, Q.; Zhang, W.; Yang, J. J.; Qian, H., Fully Hardware-Implemented Memristor Convolutional Neural Network. *Nature* **2020,** *577* (7792), 641-646.
33. Liu, Z.; Tang, J.; Gao, B.; Yao, P.; Li, X.; Liu, D.; Zhou, Y.; Qian, H.; Hong, B.; Wu, H., Neural Signal Analysis with Memristor Arrays Towards High-Efficiency Brain-Machine Interfaces. *Nature Communications* **2020,** *11* (1).
34. Kutejova, L.; Vilcakova, J.; Moucka, R.; Kazantseva, N. E.; Winkler, M.; Babayan, V., A Solvent Dispersion Method for the Preparation of Silicone Composites Filled with Carbon Nanotube. *Chemicke Listy* **2014,** *108*, S78-S85.
35. Burch, M. J.; Ievlev, A. V.; Mahady, K.; Hysmith, H.; Rack, P. D.; Belianinov, A.; Ovchinnikova, O. S., Helium Ion Microscopy for Imaging and Quantifying Porosity at the Nanoscale. *Analytical Chemistry* **2018,** *90* (2), 1370-1375.
36. Suwanratchatamanee, K.; Matsumoto, M.; Hashimoto, S., Haptic Sensing Foot System for Humanoid Robot and Ground Recognition with One-Leg Balance. *Ieee Transactions on Industrial Electronics* **2011,** *58* (8), 3174-3186.
37. Zhou, Y.; He, J.; Wang, H.; Qi, K.; Nan, N.; You, X.; Shao, W.; Wang, L.; Ding, B.; Cui, S., Highly Sensitive, Self-Powered and Wearable Electronic Skin Based on Pressure-Sensitive Nanofiber Woven Fabric Sensor. *Scientific Reports* **2017,** *7*, 12949.
38. Yin, B.; Liu, X.; Gao, H.; Fu, T.; Yao, J., Bioinspired and Bristled Microparticles for Ultrasensitive Pressure and Strain Sensors. *Nature Communications* **2018,** *9*, 5161.
39. Yang Wang, H. W., Lin Xu, Hainan Zhang, Ya Yang, Zhong Lin Wang, Hierarchically Patterned Self-Powered Sensors for Multifunctional Tactile Sensing. *Science Advances* **2020,** *6*, eabb9083.
40. Lee, J.; Kim, G.; Shin, D.-K.; Park, J., Solution-Processed Resistive Pressure Sensors Based on Sandwich Structures Using Silver Nanowires and Conductive Polymer. *Ieee Sensors Journal* **2018,** *18* (24), 9919-9924.
41. Park, H.; Jeong, Y. R.; Yun, J.; Hong, S. Y.; Jin, S.; Lee, S.-J.; Zi, G.; Ha, J. S., Stretchable Array of Highly Sensitive Pressure Sensors Consisting of Polyaniline Nanofibers and Au-Coated Polydimethylsiloxane Micropillars. *Acs Nano* **2015,** *9* (10), 9974-9985.

42. Lv, L.; Zhang, P.; Xu, T.; Qu, L., Ultrasensitive Pressure Sensor Based on an Ultralight Sparkling Graphene Block. *Acs Applied Materials & Interfaces* **2017,** *9* (27), 22885-22892.
43. Chen, X.; Liu, H.; Zheng, Y.; Zhai, Y.; Liu, X.; Liu, C.; Mi, L.; Guo, Z.; Shen, C., Highly Compressible and Robust Polyimide/Carbon Nanotube Composite Aerogel for High-Performance Wearable Pressure Sensor. *ACS Applied Materials & Interfaces* **2019,** *11*, 42594-42606.
44. Tao, L.-Q.; Zhang, K.-N.; Tian, H.; Liu, Y.; Wang, D.-Y.; Chen, Y.-Q.; Yang, Y.; Ren, T.-L., Graphene-Paper Pressure Sensor for Detecting Human Motions. *ACS Nano* **2017,** *11* (9), 8790-8795.
45. Mao, Y.; Ji, B.; Chen, G.; Hao, C.; Zhou, B.; Tian, Y., Robust and Wearable Pressure Sensor Assembled from AgNW-Coated PDMS Micropillar Sheets with High Sensitivity and Wide Detection Range. *ACS Applied Nano Materials* **2019,** *2* (5), 3196-3205.
46. Pan, L.; Chortos, A.; Yu, G.; Wang, Y.; Isaacson, S.; Allen, R.; Shi, Y.; Dauskardt, R.; Bao, Z., An Ultra-Sensitive Resistive Pressure Sensor Based on Hollow-Sphere Microstructure Induced Elasticity in Conducting Polymer Film. *Nature Communications* **2014,** *5*, 3002.
47. Zhan, Z.; Lin, R.; Tran, V.-T.; An, J.; Wei, Y.; Du, H.; Tran, T.; Lu, W., Paper/Carbon Nanotube-Based Wearable Pressure Sensor for Physiological Signal Acquisition and Soft Robotic Skin. *ACS Applied Materials & Interfaces* **2017,** *9* (43), 37921-37928.
48. Li, P.; Zhao, L.; Jiang, Z.; Yu, M.; Li, Z.; Zhou, X.; Zhao, Y., A Wearable and Sensitive Graphene-Cotton Based Pressure Sensor for Human Physiological Signals Monitoring. *Scientific Reports* **2019,** *9*, 14457.
49. Kweon, O. Y.; Lee, S. J.; Oh, J. H., Wearable High-Performance Pressure Sensors Based on Three-Dimensional Electrospun Conductive Nanofibers. *Npg Asia Materials* **2018,** *10*, 540-551.
50. Huang, Y.; Chen, Y.; Fan, X.; Luo, N.; Zhou, S.; Chen, S.-C.; Zhao, N.; Wong, C. P., Wood Derived Composites for High Sensitivity and Wide Linear-Range Pressure Sensing. *Small* **2018,** *14* (31), 1801520.
51. Luo, N.; Huang, Y.; Liu, J.; Chen, S.-C.; Wong, C. P.; Zhao, N., Hollow-Structured Graphene-Silicone-Composite-Based Piezoresistive Sensors: Decoupled Property Tuning and Bending Reliability. *Advanced Materials* **2017,** *29* (40), 1702675.
52. Lele, P. P.; Sinclair, D. C.; Weddell, G., The Reaction Time to Touch. *Journal of Physiology-London* **1954,** *123* (1), 187-203.
53. Sun, X.; Sun, J.; Li, T.; Zheng, S.; Wang, C.; Tan, W.; Zhang, J.; Liu, C.; Ma, T.; Qi, Z.; Liu, C.; Xue, N., Flexible Tactile Electronic Skin Sensor with 3D Force Detection Based on Porous CNTs/PDMS Nanocomposites. *Nano-Micro Letters* **2019,** *11* (1), 57.
54. Gu, J.; Han, J.; Liu, D.; Yu, X.; Kang, L.; Qiu, S.; Jin, H.; Li, H.; Li, Q.; Zhang, J., Solution-Processable High-Purity Semiconducting SWCNTs for Large-Area Fabrication of High-Performance Thin-Film Transistors. *Small* **2016,** *12* (36), 4993-4999.
55. Yu, X.; Liu, D.; Kang, L.; Yang, Y.; Zhang, X.; Lv, Q.; Qiu, S.; Jin, H.; Song, Q.; Zhang, J.; Li, Q., Recycling Strategy for Fabricating Low-Cost and High-Performance Carbon Nanotube TFT Devices. *Acs Applied Materials & Interfaces* **2017,** *9* (18), 15719-15726.
56. Shin, S.-H.; Ji, S.; Choi, S.; Pyo, K.-H.; An, B. W.; Park, J.; Kim, J.; Kim, J.-Y.; Lee, K.-S.; Kwon, S.-Y.; Heo, J.; Park, B.-G.; Park, J.-U., Integrated Arrays of Air-Dielectric Graphene Transistors as Transparent Active-Matrix Pressure Sensors for Wide Pressure Ranges. *Nature Communications* **2017,** *8*, 14950.
57. Takei, K.; Takahashi, T.; Ho, J. C.; Ko, H.; Gillies, A. G.; Leu, P. W.; Fearing, R. S.; Javey, A., Nanowire Active-Matrix Circuitry for Low-Voltage Macroscale Artificial Skin. *Nature Materials* **2010,** *9* (10), 821-826.
58. Kaltenbrunner, M.; Sekitani, T.; Reeder, J.; Yokota, T.; Kuribara, K.; Tokuhara, T.; Drack, M.; Schwoediauer, R.; Graz, I.; Bauer-Gogonea, S.; Bauer, S.; Someya, T., An Ultra-Lightweight Design for Imperceptible Plastic Electronics. *Nature* **2013,** *499* (7459), 458-463.
59. Someya, T.; Sekitani, T.; Iba, S.; Kato, Y.; Kawaguchi, H.; Sakurai, T., A Large-Area, Flexible Pressure Sensor Matrix with Organic Field-Effect Transistors for Artificial Skin Applications. *PNAS* **2004,** *101* (27), 9966-9970.
60. Wang, C.; Hwang, D.; Yu, Z.; Takei, K.; Park, J.; Chen, T.; Ma, B.; Javey, A., User-Interactive Electronic Skin for Instantaneous Pressure Visualization. *Nature Materials* **2013,** *12* (10), 899-904.
61. Jang, J.; Kim, H.; Ji, S.; Kim, H. J.; Kang, M. S.; Kim, T. S.; Won, J.-e.; Lee, J.-H.; Cheon, J.; Kang, K.; Im, W. B.; Park, J.-U., Mechanoluminescent, Air-Dielectric $MoS_2$ Transistors as Active-Matrix Pressure Sensors for Wide Detection Ranges from Footsteps to Cellular Motions. *Nano Letters* **2020,** *20* (1), 66-74.
62. Dahiya, R. S.; Metta, G.; Valle, M.; Sandini, G., Tactile Sensing-From Humans to Humanoids. *Ieee Transactions on Robotics* **2010,** *26* (1), 1-20.
63. Liu, Q.; Gao, B.; Yao, P.; Wu, D.; Chen, J.; Pang, Y.; Zhang, W.; Liao, Y.; Xue, C.-X.; Chen, W.-H.;

Tang, J.; Wang, Y.; Chang, M.-F.; Qian, H.; Wu, H.; Ieee, A Fully Integrated Analog ReRAM Based 78.4TOPS/W Compute-In-Memory Chip with Fully Parallel MAC Computing. *2020 IEEE International Solid- State Circuits Conference* **2020**, 500-502.

64.     Lin, W.; Wang, B.; Peng, G.; Shan, Y.; Hu, H.; Yang, Z., Skin-Inspired Piezoelectric Tactile Sensor Array with Crosstalk-Free Row Plus Column Electrodes for Spatiotemporally Distinguishing Diverse Stimuli. *Advanced Science* **2021,** *8* (3), 2002817.